\documentclass[preprint,aps,floatfix,nofootinbib,groupedaddress,superscriptaddress]{revtex4}
\usepackage{float}
\usepackage{amsfonts}
\usepackage{amssymb}
\usepackage{savesym}
\usepackage{amsmath}
\savesymbol{iint}
\usepackage{txfonts}
\restoresymbol{TXF}{iint}
\usepackage{ctable}

\usepackage{epsfig}
\usepackage{textcomp}
\usepackage[page]{appendix}
\usepackage{bm}

\begin{document}

\title{Entanglement of polar symmetric top molecules as candidate qubits}

\author{Qi Wei}
\affiliation{Department of Physics, Texas A $\&$ M University,
College Station, TX 77843, USA}
\author{Sabre Kais}
\affiliation{Department of Chemistry, Purdue University, West
Lafayette, IN 47907, USA}
\author{Bretislav Friedrich}
\affiliation{ Fritz-Haber-Institut der Max-Planck-Gesellschaft,
Faradayweg 4-6, D-14195 Berlin, Germany}
\author{ Dudley Herschbach \footnote{ Corresponding email: dherschbach@gmail.com}} \affiliation{Department of Physics,
Texas A $\&$ M University, College Station, TX 77843, USA}

\begin{abstract}
Proposals for quantum computing using rotational states of polar
molecules as qubits have previously considered only diatomic
molecules.  For these the Stark effect is second-order, so a sizable
external electric field is required to produce the requisite dipole
moments in the laboratory frame. Here we consider use of polar
symmetric top molecules.  These offer advantages resulting from a
first-order Stark effect, which renders the effective dipole moments
nearly independent of the field strength.  That permits use of much
lower external field strengths for addressing sites.  Moreover, for
a particular choice of qubits, the electric dipole interactions
become isomorphous with NMR systems for which many techniques
enhancing logic gate operations have been developed. Also inviting
is the wider chemical scope, since many symmetric top organic
molecules provide options for auxiliary storage qubits in spin and
hyperfine structure or in internal rotation states.
\end{abstract}

\maketitle

\section{Introduction}
In principle, a quantum computer can perform a variety of
calculations with exponentially fewer steps than a classical
computer \cite{Bennett,Deutsch,Feynman,Shor,Grover,Chuang}. This
prospect has fostered many proposals for means to implement a
quantum computer
\cite{Seth,DiVincenzo,Gershenfeld,Cory,Kane,Loss,Demille,wallraff,Lee,andre,yelin}.
Using arrays of trapped ultracold polar molecules is considered a
promising approach, particularly since it appears feasible to scale
up such systems to obtain large networks of coupled qubits
\cite{Demille,andre,yelin,carr,Book2009,Friedrich,Kotochigova,Micheli,Charron,kuz,ni,lics,YelinDeMille,Wei1,Wei2}.
Molecules offer a variety of long-lived internal states, often
including spin or hyperfine structure as well as rotational states.
The dipole moments available for polar molecules provide a ready
means to address and manipulate qubits encoded in rotational states
via interaction with external electric fields as well as photons.

Entanglement of qubit states, a major ingredient in quantum
computation algorithms, occurs in polar molecule arrays by
dipole-dipole interactions.    In a previous study, we examined how
the external electric field, integral to current designs for quantum
computation with polar molecules, affects both the qubit states and
the dipole-dipole interaction \cite{Wei2}. As in other work
concerned with entanglement of electric dipoles, we considered
diatomic or linear molecules, for which the Stark effect is
ordinarily second-order. Consequently, a sizable external field
($\sim$several kV/cm) is required to obtain the requisite effective
dipole moments in the laboratory frame.

In considering the operation of a key quantum logic gate (CNOT), we
evaluated a crucial parameter, $\triangle\omega$, due to the
dipole-dipole interaction.  This is the shift in the frequency for
transition between the target qubit states when the control qubit
state is changed.   For candidate diatomic molecules, under
anticipated conditions for proposed designs, $\triangle\omega$ is
very small ($\sim$20-60 kHz).   It is essential to be able to
resolve the $\triangle\omega$ shift unambiguously, but in view of
line broadening expected with a sizeable external field, whether
that will be feasible remains an open question \cite{Wei2}.

This question led us to consider polar symmetric top molecules, for
which the Stark effect is first-order in most rotational states. The
effective dipole moments are then nearly independent of the field
strength. That enables use of a much lower external field (a few
V/cm) to address and manipulate the dipoles, improving prospects for
resolving the $\triangle\omega$ shift.  The constancy of the
symmetric top effective dipole moments also makes entanglement
properties of electric dipole interactions isomorphous with those
for nuclear magnetic resonance systems.  This suggests that NMR
techniques, extensively developed for  quantum computation but
limited in application by the small size of nuclear spins and
scalability prospects \cite{Seth,Cory2,Lieven} might find congenial
applications with qubit systems comprised of polar symmetric top
molecules.

\section{EIGENSTATES FOR A POLAR SYMMETRIC TOP }

The Hamiltonian for a single trapped polar symmetric top molecule in
an external electric field may be written
\begin{equation}
{\bf H} = {\bf H}_R + {\bf H}_S + {\bf H}_T + {\bf H}_{s.q}
\end{equation}
The major term is the rotational energy
\begin{equation}
{\bf H}_R = B{\mathbf J}^2 + (A-B){\mathbf J}_z^2
\end{equation}
where $\mathbf{J}$ denotes the total rotational angular momentum and
${\mathbf J}_z$ its projection on the symmetry axis; A and B, the
rotational constants, nominally inversely proportional to the
moments of inertia about the principal axes along and perpendicular
to the symmetry axis, respectively (actually effective values
averaged over vibration and centrifugal distortion of the molecule).
The Stark energy from interaction with the external electric field
is
\begin{equation}
{\bf H}_S = -\boldsymbol{\mu}\cdot \text{\Large
\boldsymbol{$\varepsilonup$}} = -{\bf \mu}\text{\Large $\mathbf{
\varepsilonup}$}\text{cos}\theta
\end{equation}
with $\theta$ the angle between the body-fixed dipole moment $\mu$
(along the symmetry axis) and the direction of the field. The
trapping energy is
\begin{equation}
{\bf H}_T=\frac{p^2}{2m} + V_{trap}
\end{equation}
but at ultracold temperatures the translational kinetic energy
$p^2/2m$ is quite small and very nearly harmonic within the trapping
potential $V_{trap}$; thus ${\bf H}_T$ is nearly constant and for
our purposes can be omitted.   The remaining term, ${\bf H}_{s,q}$,
represents interactions arising from nuclear spins and/or quadrupole
moments; here we omit treating these, except for an important effect
of the quadrupole interaction in modifying selection rules for
transitions between qubit states.

In familiar notation, \cite{Zare,Townes} the eigenenergy for ${\bf
H}_R$ is
\begin{equation}
E_R(J,K) = BJ(J+1) + (A-B)K^2
\end{equation}
For a prolate top, $A > B$; for an oblate top, $A < B$. The Stark
energy for ${\bf H}_S$ is
\begin{equation}
E_S(J,K,M_J) = -{\bf \mu}\text{\Large $\mathbf{
\varepsilonup}$}M_JK/J(J+1)
\end{equation}
to first order. The second-order term is far smaller \cite{Stark}
(so neglected here) except for $K = 0$ or $M_J = 0$ states (which we
will not use as qubits). The corresponding eigenfunction for ${\bf
H}_R$ can be written as \cite{Zare,Townes}
\begin{equation}
|J,K,M\rangle =
(-1)^{M-K}\left[\frac{2J+1}{8\pi^2}\right]^{\frac{1}{2}}e^{i \phi M}
d_{-M -K}^J(\theta) e^{i \chi K}
\end{equation}
where $\phi$, $\theta$ and $\chi$ are the Euler angles and
$d_{-M-K}^J(\theta)$ is a Jacobi polynomial (aside from a simple
prefactor). Hence, in addition to the polar angle $\theta$ that
governs the Stark interaction, the eigenfunction depends on the
azimuthal angles $\chi$ and $\phi$ associated with, respectively,
the projections of $\mathbf{J}$ on the molecular symmetry axis and
on the {\Large\boldsymbol{$\varepsilonup$}}-field direction.

Figure \ref{Fig1} displays for the $K = 1$ sublevels of the $J = 1$
and 2 symmetric top rotational states the (a) eigenenergies $W = E_R
+ E_S$ and (b) expectation values $\langle\text{cos}\rangle =
\mu_{eff}/\mu$ for the projection of the dipole moment on the field
direction, as functions of $\mu${\Large $\varepsilonup$}/$B$. The
dependence on $\mu${\Large $\varepsilonup$}/$B$ differs markedly
from a similar plot for a diatomic molecule (for which $K = 0$; {\it
cf.} Fig. 1 of ref. \cite{Wei2}); there the effective dipole moments
are field-dependent and vanish at zero-field. For symmetric top
qubit states, to take advantage of the first-order Stark effect, we
consider only $K \not= 0$ and $M_J \not= 0$ states. For such states,
the effective dipole moments,
\begin{equation}
\mu_{eff} = -\partial E_S/\partial \text{\Large $\mathbf{
\varepsilonup}$} = \mu M_JK/J(J+1)
\end{equation}
are just constants independent of the field (except at unusually
high fields, where higher order terms become important
\cite{Stark}). According as $\mu_{eff}$ is positive or negative, the
Stark energy drops or climbs as the field strength grows, so   the
molecular states are termed high field seeking (HFS) or low field
seeking LFS), respectively.

\subsection{ Choice of qubit states }

\begin{figure}[t]
\begin{center}
\includegraphics[width=0.90\textwidth]{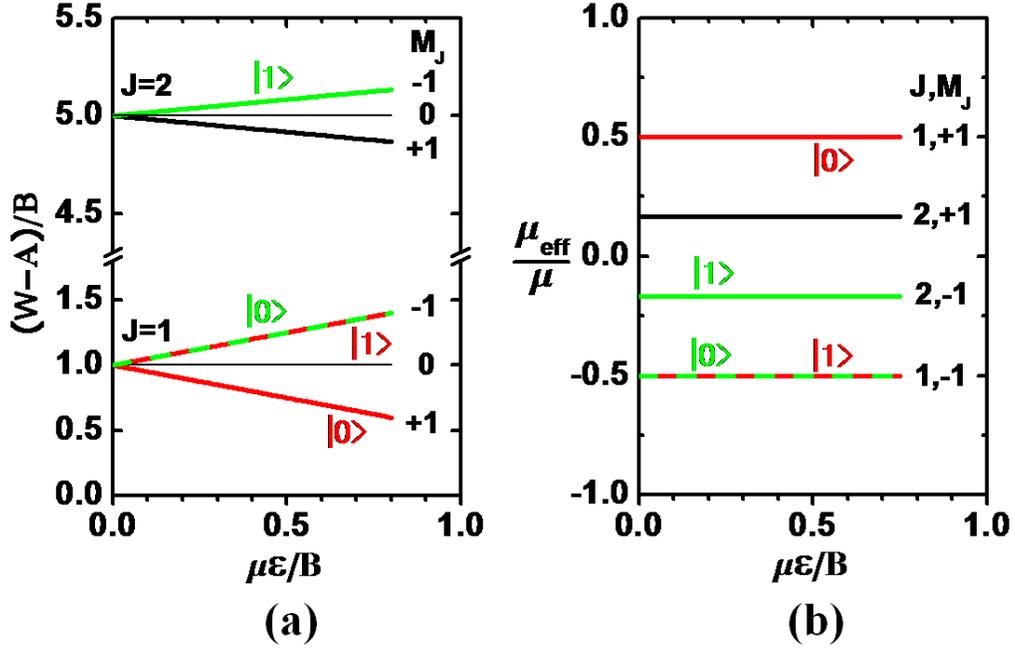}
\end{center}
\caption{(Color online)Stark states for a polar symmetric top
molecule, as functions of $\mu${\Large $\varepsilonup$}$/B$.   (a)
Eigenenergies for $M_J = 0$ and $\pm 1$ components of $K = 1$ levels
for $J = 1$ and $J = 2$ states and (b) corresponding expectation
values that determine effective dipole moments, $\mu_{eff} =
\langle\text{cos}\rangle$. States used as basis qubits are labeled
$|0\rangle$ and $|1\rangle$: type I (green) are $M_J = -1$ for $J =
2$ and type II (red) are $M_J = +1$ and -1 for $J = 1$.  By virtue
of the ordinate scale used, (a) as well as (b) applies to any
symmetric top molecule (treated as rigid, without fine or hyperfine
structure). } \label{Fig1}
\end{figure}

We consider two qualitatively distinct choices for qubit states,
designated I and II. The orthodox choice, type I, is exemplified by
\begin{equation}
|0\rangle = |J=1,K=1,M_J=-1\rangle \;\;\;\;\; \text{and} \;\;\;\;\;
|1\rangle = |J=2,K=1,M_J=-1\rangle
\end{equation}
For this choice (green curves in Fig. \ref{Fig1}), radiation induced
transitions between the qubits are fully allowed, in accord with the
familiar selection rules, $\triangle J = 0, \pm 1$; $\triangle K =
0$; $\triangle M = 0, \pm 1$ \cite{Townes}. Also, both the
$|0\rangle$ and $|1\rangle$ qubit states are LFS, thereby
facilitating trapping by either DC or AC fields or an optical
lattice \cite{Meerakker}. The corresponding eigenenergies, $E_R +
E_S$, are
\begin{equation}
W_0 = A + B + \frac{\mu \text{\Large $\mathbf{ \varepsilonup}$}}{2}
\;\;\;\;\; \text{and} \;\;\;\;\; W_1 = A + 5B + \frac{\mu
\text{\Large $\mathbf{ \varepsilonup}$}}{6}
\end{equation}
and the $\text{cos}\theta$ matrix elements are
\begin{equation}
C_0 = \langle 0|\text{cos}\theta|0\rangle = -\frac{1}{2}, \;\;\;\;\;
\;\;\;\;\; C_1 = \langle 1|\text{cos}\theta|1\rangle =
-\frac{1}{6},\;\;\;\;\; \;\;\;\;\; C_X = \langle
0|\text{cos}\theta|1\rangle = \frac{\sqrt{15}}{10} \label{C0C1}
\end{equation}

We are particularly interested in an unorthodox choice, type II (red
curves in Fig. \ref{Fig1}). For this, the qubit states are
\begin{equation}
|0\rangle = |J=1,K=1,M_J=+1\rangle \;\;\;\;\; \text{and} \;\;\;\;\;
|1\rangle = |J=1,K=1,M_J=-1\rangle
\end{equation}
The eigenenergies are degenerate at zero-field but for $\text{\Large
$\mathbf{ \varepsilonup}$} > 0$ split apart strongly and linearly,
\begin{equation}
W_0 = A + B - \frac{\mu \text{\Large $\mathbf{ \varepsilonup}$}}{2}
\;\;\;\;\; \text{and} \;\;\;\;\; W_1 = A + B + \frac{\mu
\text{\Large $\mathbf{ \varepsilonup}$}}{2}
\end{equation}
and the $\text{cos}\theta$ matrix elements are
\begin{equation}
C_0 = \langle 0|\text{cos}\theta|0\rangle = \frac{1}{2}, \;\;\;\;\;
\;\;\;\;\; C_1 = \langle 1|\text{cos}\theta|1\rangle =
-\frac{1}{2},\;\;\;\;\; \;\;\;\;\; C_X = \langle
0|\text{cos}\theta|1\rangle = 0 \label{C0C1}
\end{equation}
These type II qubits render the effective dipole moments constant
and equal in magnitude but opposite in sign.  However, type II
qubits require further specification.   As initially defined in
Eq.(12), the transition $|0\rangle \leftrightarrow |1\rangle$
between the qubits requires $\triangle M_J = \pm 2$.  Thus, it is
not allowed as a one-photon electric dipole transition (the
transition cosine, $C_X = 0$).  It is allowed as a two-photon
transition (using the $J = 1$, $K = 1$, $M_J = 0$ state as
intermediate). Another remedy, simpler to implement, is to use a
molecule that contains a nuclear quadrupole moment.  Even a small
quadrupole coupling constant typically introduces sufficient mixing
of Stark states to make $\triangle M_J = \pm 2$ transitions become
prominent in microwave or radiofrequency spectra \cite{Klemperer}.
In accord with theory \cite{Townes,Coester}, in the next subsection
we show that modifying the type II qubit choice to exploit the
quadrupole hyperfine structure renders $C_X \not= 0$, enabling
$|0\rangle \leftrightarrow |1\rangle$ to occur as a one-photon
transition.

In another contrast with type I, for type II qubits $|0\rangle$ is
HFS while $|0\rangle$ is LFS. That is also often the case for qubit
states considered for diatomic molecules, and is not regarded as a
serious handicap \cite{Meerakker}.  Although HFS states are harder
to trap, both HFS and LFS can be captured simultaneously in an AC
trap or an optical lattice \cite{Meerakker}.

\subsection{Quadrupole perturbation of Stark states}

For simplicity, we consider symmetric top molecules having only one
atom with a nuclear quadrupole moment, with that atom located on the
symmetry axis.   We also treat explicitly only cases in which the
nuclear spin $I = 1$ for that atom, and the quadrupole interaction
is much smaller than the Stark energy.   The $CH_3CN$ molecule
\cite{Kukolich} is a prototypical case: for the $^{14}N$ nucleus
(spin $I = 1$), the quadrupole coupling constant is $eqQ = -4.22
\;\text{MHz}$. For conditions in prospect for a quantum computer,
usually $\mu${\Large $\varepsilonup$}$\; > 100 \;\text{MHz}$. A
first-order perturbation treatment, referred to as the "strong-field
approximation" \cite{Townes,Coester}, and governed by the ratio
$eqQ/\mu${\Large $\varepsilonup$}, hence is appropriate for this
example and many others.

When set-up in the usual $|JKM_JIM_I\rangle$ basis, with $M_I$ the
projection of the nuclear spin on the {\Large $\varepsilonup$}-field
direction, the Hamiltonian matrix, ${\bf H}_R + {\bf H}_S + {\bf
H}_Q$, is diagonal in $J$, $K$, and $I$.  The ${\bf H}_R$ and ${\bf
H}_S$ portions are also diagonal in $M_J$ and $M_I$ whereas ${\bf
H}_Q$ has off-diagonal elements which connect $M_J$ and $M_I$ states
differing by up to two units. In consequence of the resulting
mixing, neither $M_J$ nor $M_I$ is a "good" quantum number. Their
sum, $M_J + M_I$ remains good, however, since the total angular
momentum along the field must be constant. Accordingly, we modify
our choices for the $|0\rangle$ and $|1\rangle$ qubits of Eqs.(9)
and (12), that involve $M_J = \pm 1$, to specify them further as
particular hyperfine components with $M_J + M_I = 0$. In Appendix A
we evaluate the contributions from ${\bf H}_Q$ to the qubit
eigenenergies and cosine matrix elements.

In first-order, the quadrupole interaction simply adds to the qubit
eigenvalues of Eq. (10) or (13) a diagonal term given by
\begin{equation}
E_Q = eqQ/40 \;\;\;\;\;\text{or}\;\;\;\;\; eqQ/56
\end{equation}
for $J = 1$ or $J = 2$, respectively.

The cosine matrix elements of Eqs.(11) and (14) are augmented by
terms involving $w = |eqQ|/\mu${\Large $\varepsilonup$}, given in
Table I. Since typically $w < 10^{-2}$, these contributions are
insignificant for type I qubits, and for the $C_0$ or $C_1$ elements
for type II qubits, but of major importance in the $C_X$ transition
element for type II, which would otherwise be zero. Even when $C_X$
is very small, conventional power levels suffice to make transitions
facile between the $M_J = \pm 1$ Stark components \cite{Klemperer}.

\begin{table}[H]
\begin{minipage}{13cm}
\centering \caption{Cosine matrix elements for symmetric top
qubits$^a$.}
\begin{tabular}{ccc}
\hline\hline
      \;&\;  Type I qubits  \;&\; Type II qubits \\
\hline\let\thefootnote\relax\footnotetext{$^a$Terms in $w =
|eqQ|/\mu${\Large $\varepsilonup$} are contributions from quadrupole
coupling.  These were fitted to results of numerical calculations
(see Appendix A) extending over the range $w < 1$.}
$C_0$ \;\;\;\;\;&\;\;\;\;\;  $-1/2 - 0.00168w + 0.0418w^2$          \;\;\;\;\;&\;\;\;\;\;  $1/2 - 0.00347w - 0.0213w^2$  \\
$C_1$ \;\;\;\;\;&\;\;\;\;\;  $-1/6 + 0.00526w + 0.0218w^2$          \;\;\;\;\;&\;\;\;\;\;  $-1/2 - 0.00168w + 0.0418w^2$  \\
$C_X$ \;\;\;\;\;&\;\;\;\;\;  $\sqrt{15}/10 - 0.00658w - 0.0437w^2$  \;\;\;\;\;&\;\;\;\;\;  $0 + 0.153w - 0.0108w^2$  \\
\hline \hline
\end{tabular}
\end{minipage} \label{table-fitC}
\end{table}

\section{TWO INTERACTING DIPOLES}

Adding a second trapped polar symmetric top, identical to the first
but a distance $r_{12}$ apart, introduces the dipole-dipole coupling
interaction,
\begin{equation}
V_{d-d} =\frac{\bm{\mu}_1\cdot\bm{\mu}_2-3(\bm{\mu}_1 \cdot {\bf
n})(\bm{\mu}_2 \cdot {\bf n})}{|{\bf r}_1-{\bf r}_2|^3}
\label{coupling}
\end{equation}
Here {\bf n} denotes a unit vector along ${\bf r}_{12}$.  In the
presence of an external field, it becomes appropriate to express
$V_{d-d}$ in terms of angles related to the field direction
(Appendix A in ref. \cite{Wei2}). The result after averaging over
azimuthal angles reduces to
\begin{equation}
V_{d-d}=\Omega(1-3\text{cos}^2\alpha)\text{cos}\theta_1\text{cos}\theta_2
\label{coupling2}
\end{equation}
where $\Omega=\mu^2/r_{12}^3$, the angle $\alpha$ is between the
${\bf r}_{12}$ vector and the field direction and polar angles
$\theta_1$ and $\theta_2$ are between the ${\bm \mu}_1$ and ${\bm
\mu}_2$ dipoles and the field direction.

When set up in a basis of the qubit states (either type I or II) for
the pair of molecules,
$\{|00\rangle,|01\rangle,|10\rangle,|11\rangle\}$, the $H_R + H_S$
portion of the Hamiltonian takes the form
\begin{equation}
\left(\begin{array}{cccc}
W_0+W_0^{\prime}  & 0       & 0       & 0   \\
 0    & W_0+W_1^{\prime} & 0       & 0   \\
 0    & 0       & W_1+W_0^{\prime} & 0  \\
 0    & 0       & 0       & W_1+W_1^{\prime}
\end{array} \right)
\label{Hs-matrix}
\end{equation}
and the $V_{d-d}$ portion is
\begin{equation}
\Omega_\alpha \left(\begin{array}{cccc}
C_0C_0^{\prime}  & C_0C_X^{\prime} & C_XC_0^{\prime} & C_XC_X^{\prime}   \\
 C_0C_X^{\prime} & C_0C_1^{\prime} & C_XC_X^{\prime} & C_XC_1^{\prime}   \\
 C_XC_0^{\prime} & C_XC_X^{\prime} & C_1C_0^{\prime} & C_1C_X^{\prime}  \\
 C_XC_X^{\prime} & C_XC_1^{\prime} & C_1C_X^{\prime} & C_1C_1^{\prime}
\end{array} \right)
\label{Hdd-matrix1}
\end{equation}
where $\Omega_\alpha = \Omega(1-3\text{cos}^2\alpha)$.The primes
attached to quantities for the second dipole indicate that the
external field magnitude will differ at its site; that is necessary
for addressing the sites and to ensure that the qubit states
$|01\rangle$ and $|10\rangle$ differ in energy.

\subsection{Evaluating entanglement of eigenstates}

The form of the Hamiltonian in Eqs. (18) and (19) is identical to
that for two polar diatomic molecules, treated in ref. \cite{Wei2}.
Thus, we follow the same procedures in evaluating eigenstate
properties and entanglement for symmetric tops, merely introducing
the appropriate matrix elements for qubits of types I and II  (as
specified in Sec IIA).   We again use unitless reduced variables,
$x$ =  $\mu${\Large $\varepsilonup$}$/B$ and $y =
\Omega_{\alpha}/B$; in terms of customary units, these are given by
\begin{equation}
x = \mu \text{\Large $\mathbf{ \varepsilonup}$}/B = 504
\mu(\text{Debye})\text{\Large $\mathbf{
\varepsilonup}$}(\text{kV/cm})/B(\text{MHz})
\end{equation}
\begin{equation}
y = \Omega_\alpha/B = 1.51 \times 10^{-4} \mu^2
(\text{Debye})/r^3(\mu m)/B (\text{MHz})
\end{equation}
Likewise, we use $z = eqQ/B$ for quadrupole coupling terms.  The
pertinent ranges are  $x < 1$,  $y < 10^{-5}$, and $|z| < 5 \times
10^{-3}$ for candidate symmetric tops (with dipole moments $\mu < 4$
D, quadrupole coupling $|eqQ| < 10$ MHz,  and rotational constants
$B > 2000$ MHz) under conditions deemed practical  for prospective
quantum computer designs (field strengths {\Large $\varepsilonup$}
$< 1$ kV/cm, intermolecular spacings $r \sim 0.5 \mu m$).  Unless
otherwise noted, we take $\alpha = 90^o$.  In the pertinent regime,
the dependence on $x$, $y$, and $z$ of the eigenenergies is simply
linear in all three variables.

Another key variable is  $\triangle x = x' - x$, specified by the
difference in the field strength at adjacent qubit sites.   As the
site addresses are provided by observing the one-qubit transition,
$|0\rangle \leftrightarrow |1\rangle$, the size of $\triangle x$
must be large enough to produce a clearly resolvable Stark shift
between the sites.   Yet $\triangle x$ must not exceed $X_R/N$,
where $N$ is the number of sites and $X_R$ the range in $x$ of field
strengths considered feasible. To benefit from keeping the field
strength relatively low, we take $X_R \sim 1$; then to accommodate
$N$ sites requires $\triangle x < X_R/N$.  At least for exploratory
calculations for up to $N \sim 10^3$, we consider $10^{-4} <
\triangle x < 10^{-2}$ appropriate.

Tables II and III exhibit properties, for qubit types I and II,
respectively, of the four eigenstates of the two-dipole system,
listed in order of increasing energy ($i = 1 \rightarrow 4$).   The
eigenvalues are obtained as simple explicit functions of $x$, $x'$,
$y$, $z$, applicable to any polar symmetric top molecule and
conditions within the pertinent regime specified above.   Also
indicated, in order of magnitude only, are quantities that express
the extent of entanglement among the qubit basis states, but must be
evaluated by numerical means.  Entanglement is exhibited most
directly in the coefficients with which the qubit basis states
appear in the eigenfunctions,
\begin{equation}
\Psi_i = a_i|00\rangle + b_i|01\rangle + c_i|10\rangle +
d_i|11\rangle
\end{equation}
In Appendix B, we give somewhat cumbersome formulas for these
coefficients in terms of $x$, $x'$, $y$, $z$. Tables II and III show
just orders of magnitude, evaluated for $CH_3CN$, under conditions
specified in Table IV.  This is done to illustrate most simply a
major point: In the pertinent range,  the entanglement is so feeble
that the successive eigenfunctions $\Psi_i$ differ only slightly
from the respective basis qubits,
$\{|00\rangle,|01\rangle,|10\rangle,|11\rangle\}$; there is little
admixture with other qubits.

\begin{table}[H]
\begin{minipage}{17cm}
\centering \caption{Eigenproperties for $N = 2$ symmetric top
dipoles, type I qubits$^a$.}
\begin{tabular}{cccccccc}
\hline\hline
$i$  \;&\;  $(E_i-2A)/B$  \;&\; $\Psi_i$  \;&\;  $|00\rangle$ \;&\; $|01\rangle$  \;&\;  $|10\rangle$  \;&\;  $|11\rangle$ \;&\; $C_{12}$ \\
\hline\let\thefootnote\relax\footnotetext{$^a$Here $x = \mu${\Large
$\varepsilonup$}$/B = 0.0107$, $y = \Omega_\alpha/B = 2 \times
10^{-6}$, $z = eqQ/B = 5 \times 10^{-4}$, $\triangle x = x' - x =
10^{-3}$.}
1\;&\; $2+\frac{x}{2}+\frac{x'}{2}+\frac{y}{4}+\frac{z}{20}$   \;&\;  \;&\; $1-O(10^{-15})$ \;&\; $+O(10^{-8})$  \;&\; $+O(10^{-8})$  \;&\; $-O(10^{-8})$   \;&\; $O(10^{-8})$\\
2\;&\; $6+\frac{x}{2}+\frac{x'}{6}+\frac{y}{12}+\frac{3z}{70}$ \;&\;  \;&\; $-O(10^{-8})$   \;&\; $1-O(10^{-7})$ \;&\; $-0.0009$      \;&\; $+O(10^{-8})$   \;&\; $0.0018$    \\
3\;&\; $6+\frac{x'}{2}+\frac{x}{6}+\frac{y}{12}+\frac{3z}{70}$ \;&\;  \;&\; $-O(10^{-8})$   \;&\; $-0.0009$      \;&\; $1-O(10^{-7})$ \;&\; $+O(10^{-8})$   \;&\; $0.0018$    \\
4\;&\; $10+\frac{x}{6}+\frac{x'}{6}+\frac{y}{36}+\frac{z}{28}$ \;&\;  \;&\; $+O(10^{-8})$   \;&\; $-O(10^{-8})$  \;&\; $-O(10^{-8})$  \;&\; $1-O(10^{-15})$ \;&\; $O(10^{-8})$\\
\hline \hline
\end{tabular}
\end{minipage} \label{table-typeI}
\end{table}

\begin{table}[H]
\begin{minipage}{17cm}
\centering\caption{Eigenproperties for $N = 2$ symmetric top
dipoles, type II qubits$^a$.}
\begin{tabular}{cccccccc}
\hline\hline
$i$  \;&\;  $(E_i-2A)/B$  \;&\; $\Psi_i$  \;&\;  $|00\rangle$ \;&\; $|01\rangle$  \;&\;  $|10\rangle$  \;&\;  $|11\rangle$ \;&\; $C_{12}$ \\
\hline\let\thefootnote\relax\footnotetext{$^a$Here $x = \mu${\Large
$\varepsilonup$}$/B = 0.0107$, $y = \Omega_\alpha/B = 2 \times
10^{-6}$, $z = eqQ/B = 5 \times 10^{-4}$, $\triangle x = x' - x =
10^{-3}$.}
1\;&\; $2-\frac{x}{2}-\frac{x'}{2}+\frac{y}{4}+\frac{z}{20}$   \;&\;  \;&\; $1-O(10^{-17})$ \;&\; $-O(10^{-9})$  \;&\; $-O(10^{-9})$  \;&\; $-O(10^{-12})$   \;&\; $O(10^{-12})$\\
2\;&\; $2+\frac{x}{2}-\frac{x'}{2}-\frac{y}{4}+\frac{z}{20}$ \;&\;  \;&\; $+O(10^{-9})$   \;&\; $1-O(10^{-17})$  \;&\; $-O(10^{-10})$  \;&\; $+O(10^{-9})$   \;&\; $O(10^{-9})$    \\
3\;&\; $2+\frac{x'}{2}-\frac{x}{2}-\frac{y}{4}+\frac{z}{20}$ \;&\;  \;&\; $+O(10^{-9})$   \;&\; $+O(10^{-10})$   \;&\; $1-O(10^{-17})$ \;&\; $+O(10^{-9})$   \;&\; $O(10^{-9})$    \\
4\;&\; $2+\frac{x}{2}+\frac{x'}{2}+\frac{y}{4}+\frac{z}{20}$ \;&\;  \;&\; $+O(10^{-12})$   \;&\; $-O(10^{-9})$   \;&\; $-O(10^{-9})$  \;&\; $1-O(10^{-17})$  \;&\; $O(10^{-12})$\\
\hline \hline
\end{tabular}
\end{minipage} \label{table-typeII}
\end{table}

\begin{table}[H]
\centering \caption{Parameters for $CH_3CN$ molecule.}
\begin{minipage}{9cm}
\begin{tabular}{ccc}
\hline\hline
\multicolumn{2}{c}{Properties}      \;\;\;\;\;\;&\;\;\;\;\;\; Reduced variables\footnote{For "pertinent" conditions, {\Large $\varepsilonup$} = 500 V/cm, $r = 0.5 \mu m$; See Eqs (20) and (21).} \\
\hline
$\mu$ \;&\;\; 3.92 D                 \;\;\;\;\;\;&\;\;\;\;\;\;  $x = \mu${\Large $\varepsilonup$}$/B = 0.0107$  \\
$B$ \;&\;\; 9198.8 MHz               \;\;\;\;\;\;&\;\;\;\;\;\;  $\triangle x = \mu$({\Large $\varepsilonup$}$'$-{\Large $\varepsilonup$})$/B = 10^{-3}$  \\
$eqQ$ \;&\;\; -4.22 MHz               \;\;\;\;\;\;&\;\;\;\;\;\;  $y = \Omega_\alpha/B = 2 \times 10^{-6}$  \\
$\mu${\Large $\varepsilonup$} \;&\;\; 988 MHz   \;\;\;\;\;\;&\;\;\;\;\;\;  $z = eqQ/B = 4.6 \times 10^{-4}$  \\
$\Omega_\alpha$ \;&\;\; 18.5 kHz     \;\;\;\;\;\;&\;\;\;\;\;\;  $w = |eqQ|/\mu${\Large $\varepsilonup$} = $4.3 \times 10^{-3}$  \\
\hline \hline
\end{tabular}
\end{minipage} \label{table-CH3CN}
\end{table}

\subsection{Pairwise concurrence of eigenstates}

A quantitative measure of entanglement is provided by the pairwise
concurrence function, $C_{12}$, which becomes unity when
entanglement is maximal and zero when it is entirely lacking. The
general prescription for evaluating $C_{12}$ involves somewhat
arcane manipulations of the density matrix \cite{Wootters}. However,
it becomes simple here as the entanglement arises entirely from
off-diagonal terms in the $V_{d-d}$ matrix of Eq.(19).   These terms
are small, since they are all proportional to $y$, which is $<
10^{-5}$. Otherwise the off-diagonal terms contain either $C_X$, or
$C_X^2$, factors essentially independent of $x$ or $x'$; for type I
qubits, $C_X \sim 0.4$ and for type II qubits $C_X < 10^{-3}$.
Accordingly, as seen in Tables II and III, the ground eigenstate,
$\Psi_1$, and the highest excited eigenstate, $\Psi_4$, are almost
solely composed of the basis qubits $|00\rangle$ and $|11\rangle$,
respectively, especially for type II. In terms of the coefficients
in Eq.(22), in this case $C_{12}$ is to good approximation just
$2d_1$ or $2a_4$, for eigenstates 1 and 4, respectively.  Thus, for
eigenstates 1 and 4, we find
\begin{equation}
C_{12} = K(x,x')[\Omega_\alpha/B]
\end{equation}
with weak dependence on $x$, given by
\begin{equation}
K(x) =  0.03752 + 0.00312x + 0.00029x^2
\end{equation}
and the  dependence on $x'$ is well represented by $K(x,x') =
\left[K(x)K(x')\right]^{1/2}$ when $\triangle x = x' - x < 10^{-2}$.
The concurrence for a pair of polar diatomic molecules \cite{Wei2}
has this same form (for small $\Omega_\alpha/B$), but the
second-order Stark effect makes the $K(x)$ coefficient much larger
($> 0.12$ for $x < 1$).

The $C_{12}$ function becomes more interesting for the middle
eigenstates, $\Psi_2$ and $\Psi_3$. As seen in Tables II and III,
for the conditions we refer to as "pertinent" these eigenstates are
essentially just the $|01\rangle$ and $|10\rangle$ basis qubits,
respectively. However, if $\triangle x \rightarrow 0$, the
eigenenergies $E_2$ and $E_3$ become the same.  In that limit, even
very small $y$ can produce strong entanglement of the $|01\rangle$
and $|10\rangle$ qubits. Figure 2 illustrates how $C_{12}$ varies as
$\triangle x$ is scanned over a range from well below to well above
$y$; at least in principle that can be done by adjusting the {\Large
$\varepsilonup$}-field and$/$or the spacing of the dipoles. The
curve shown is given by
\begin{equation}
C_{12} = 2|\alpha_{\pm}|/(1+\alpha_\pm^2)
\end{equation}
with
\begin{equation}
\alpha_\pm =
\frac{(E_3-E_2)\pm\left[(E_3-E_2)^2+4\triangle^2\right]^{1/2}}{2\triangle}
\end{equation}
where $\triangle = C_X^2\Omega_\alpha$. This formula for $C_{12}$
results from omitting all off-diagonal terms in the $V_{d-d}$ matrix
except the pair that couple $|01\rangle$ and $|10\rangle$ along the
antidiagonal. The eigenstates then become $\Psi_2 = \Psi_+$ and
$\Psi_3 = \Psi_-$, with
\begin{equation}
\Psi_\pm =
\frac{|10\rangle-\alpha_\pm|01\rangle}{\sqrt{1+\alpha_\pm^2}}
\end{equation}
In the limit $E_3 - E_2 \ll \triangle$ (i.e, $\triangle x \ll y$),
where $\alpha_\pm \rightarrow \pm 1$ and $C_{12} \rightarrow 1$, the
eigenfunctions become maximally entangled states, termed Bell
states. Figure 2 also displays points obtained from numerical
diagonalization of the Hamiltonian with all elements included in the
$V_{d-d}$ matrix.   For both type I (green points) and type II (red
points), the numerical results agree very closely with the formula
given in Eq.(25).   It is a striking demonstration of the extent to
which matrix elements that connect almost degenerate levels generate
entanglement.

\subsection{Inducing large entanglement via resonant pulses}

Under the ultracold conditions needed to localize trapped molecules
in the qubit sites, the two-dipole system is in its ground
eigenstate, $\Psi_1 \sim |00\rangle$, wherein the entanglement is
very small. However, the large entanglement often needed for quantum
computing can be induced dynamically via resonant pulses to higher
eigenstates \cite{Jones2,Thanks}.  Several procedures have been
presented for accomplishing this to use polar molecules in operating
quantum logic gates
\cite{andre,yelin,Charron,kuz,lics,YelinDeMille,Shioya,Mishima,Chen,Reina,Sugny}.
Here we consider just a rudimentary version, exemplified with the
CNOT gate, since our chief aim is to compare and contrast the
symmetric top qubits of types I and II with the diatomic case
treated in ref. \cite{Wei2}.

Figures 3 and 4 give schematic diagrams, analogous to Fig. 10 of
ref. \cite{Wei2}, depicting available transitions among the
two-dipole eigenstates. Table V lists the corresponding transition
frequencies. In contrast to type I, for type II qubits the
contributions from both the rotational constants and quadrupole
coupling cancel out, hence the transition frequencies depend only on
the Stark energy shifts and dipole-dipole interaction.   Since the
entanglement is so feeble for the eigenstates, as seen in Tables II
and III, for a heuristic description we may speak as if the
transitions simply occur between the unperturbed basis qubits.  A
typical procedure applies a $\pi/2$ pulse resonant with the
transition frequency $\omega_1$ to transfer population from the
ground eigenstate $|00\rangle$ to the excited state $|01\rangle$,
thereby putting the system into the state $2^{-1/2}(|00\rangle +
|01\rangle)$. Then a $\pi$ pulse resonant with the transition
$\omega_2$ between $|01\rangle$ and $|11\rangle$ will put the system
into the state $2^{-1/2}(|00\rangle + |11\rangle)$, which is a
completely entangled Bell state. The same process can be done
applying a $\pi/2$ pulse to $\omega_3$, followed by a $\pi$ pulse to
$\omega_4$.

\begin{table}[H]
\begin{minipage}{12cm}
\centering \caption{Transition frequencies between eigenstates of
two dipoles$^a$.}
\begin{tabular}{ccc}
\hline\hline
                      \;\;\;\;\;\;\;\;\;&\;\;\;\;\;\;\;\;\;  Type I qubits        \;\;\;\;\;\;\;\;\;&\;\;\;\;\;\;\;\;\; Type II qubits \\
\hline\let\thefootnote\relax\footnotetext{$^a$Here $x = \mu${\Large
$\varepsilonup$}$/B$, $y = \Omega_\alpha/B$, $z = eqQ/B$.}
$\omega_1/B$         \;\;\;\;\;\;\;\;\;&\;\;\;\;\;\;\;\;\;  $4-x'/3-y/6-z/140$   \;\;\;\;\;\;\;\;\;&\;\;\;\;\;\;\;\;\;  $x-y/2$  \\
$\omega_2/B$         \;\;\;\;\;\;\;\;\;&\;\;\;\;\;\;\;\;\;  $4-x/3-y/18-z/140$   \;\;\;\;\;\;\;\;\;&\;\;\;\;\;\;\;\;\;  $x'+y/2$  \\
$\omega_3/B$         \;\;\;\;\;\;\;\;\;&\;\;\;\;\;\;\;\;\;  $4-x/3-y/6-z/140$    \;\;\;\;\;\;\;\;\;&\;\;\;\;\;\;\;\;\;  $x'-y/2$  \\
$\omega_4/B$         \;\;\;\;\;\;\;\;\;&\;\;\;\;\;\;\;\;\;  $4-x'/3-y/18-z/140$  \;\;\;\;\;\;\;\;\;&\;\;\;\;\;\;\;\;\;  $x+y/2$  \\
$\triangle\omega/B$  \;\;\;\;\;\;\;\;\;&\;\;\;\;\;\;\;\;\;  $y/9$                \;\;\;\;\;\;\;\;\;&\;\;\;\;\;\;\;\;\;  $y$  \\
\hline \hline
\end{tabular}
\end{minipage} \label{table-frequencies}
\end{table}

To carry out such procedures, the transition frequencies need to be
unambiguously resolved from each other.   As evident in Table V, for
both type I and II qubits, $\omega_1$ can be resolved from
$\omega_2$ and $\omega_3$ from $\omega_4$ simply by adjusting the
difference in external field strengths, $\triangle x = x' - x$. In
frequency units, a Stark shift of $\triangle x = 10^{-3}$ for
$CH_3CN$ is 3 MHz for type I qubits and 9 MHz for type II. The
relative difference is far more in favor of type II, because
$\omega_1 = 35,869$ MHz for type I whereas it is only 988 MHz for
type II. However, for either type such differences are easily
resolvable in conventional microwave and radiofrequency
spectroscopy.

Resolving $\omega_1$ from $\omega_4$ and $\omega_2$ from $\omega_3$
presents an experimental challenge.  The frequency difference is
governed simply by the dipole-dipole interaction, since
\begin{equation}
\triangle\omega = \omega_4 - \omega_1 = \omega_2 - \omega_3 =
\Omega_{\alpha}(C_1 - C_0)(C'_1 - C'_0)
\end{equation}
The $\triangle\omega$ shift is the essential feature of a CNOT gate:
$\omega_3$ transfers the target qubit on dipole 1 from $|0\rangle$
to $|1\rangle$ when the control qubit on dipole 2 is in $|0\rangle$,
whereas $\omega_2 = \omega_3 - \triangle\omega$ transfers the target
from $|0\rangle$ to $|1\rangle$ when the control is in $|1\rangle$.
For $\omega_1$ and $\omega_4$ the roles of target and control sites
are exchanged. Unlike the diatomic case \cite{Wei2}, for symmetric
tops the cosine elements are nearly independent of the external
field in the pertinent regime, except via the minor quadrupole terms
included in Table I. Thus,
\begin{equation}
\triangle\omega = \Omega_\alpha/9 \;\;\;\;\;\text{for type I
and}\;\;\;\;\; \triangle\omega = \Omega_\alpha\;\;\;\;\;\text{for
type II}
\end{equation}
Here the significant advantage of type II occurs because both $C_0$
and $C_1$ are large and of opposite sign. In frequency units, for
$CH_3CN$ the $\triangle\omega$ shift is only 2 kHz for type I and 18
kHz for type II. Again, the relative difference greatly favors type
II, since $\triangle\omega/\omega_1$ is more than a hundredfold
larger than for type I.

As compared with candidate polar diatomic molecules \cite{Wei2}, we
expect prospects for resolving $\triangle\omega$ for symmetric tops
are improved in two ways:  (1) The first-order Stark effect enables
use of a much less strong external field. That should reduce line
broadening caused by nonuniformity and fringing of the electric
field.  (2) The choice of Stark components for type II qubits lowers
the transition frequencies between qubit states down to the
radiofrequency range (often factors of 30-50 lower than transitions
between rotational states, which occur in the microwave range).   In
molecular beam spectra, collision free but without trapping in an
optical lattice, line widths are typically much smaller in the rf
region; e.g., 2 kHz or below for $\triangle J = 0$,  $\triangle M_J
= \pm 1$ transitions \cite{Klemperer}. The effect of the optical
lattice on line widths is uncertain. It may introduce broadening via
motional shifts, which are strongly dependent on the well depths
required for trapping \cite{Barker}. Such shifts have been avoided
for ultracold atoms by use of "magic" optical trapping conditions
\cite{Zelevinsky}, but there might be less scope to do that for
molecules.   As yet, no line width data have been reported for
ultracold molecules trapped in an optical lattice and subject to a
sizable electric field. Thus, although less problematic for type II
symmetric top qubits, the feasibility of resolving the
$\triangle\omega$ shift remains an open question.

\subsection{Comparison with NMR}

A motivation for considering symmetric top type II qubits is the
resemblance to spin-1/2 NMR, which has been extensively analyzed in
the context of quantum computation
\cite{Cory,Cory2,Cory3,Vandersypen,Price,Price2}. The resemblance
stems from the unorthodox choice of $\pm M_J$ Stark components for
type II qubits. That renders the effective qubit dipole moments,
$\mu_{eff} = \mu\langle\text{cos}\theta\rangle$, which are
essentially independent of the external field, equal in size but
opposite in spatial orientation.   There are further similarities.
For the generic $N = 2$ case, the corresponding Hamiltonian for NMR
resembles our Eqs. (18) plus (19), except for omission of the
rotational energy.  The molecular dipoles are replaced by nuclear
spins, the Stark field by a Zeeman field, and the dipole-dipole
interaction by spin-spin coupling.   Thereby our $\Omega_\alpha$ is
replaced by $J_{12}$, the spin-spin coupling parameter.  Since the
Zeeman energy terms are much larger than the spin-spin coupling, the
equivalent of our $V_{d-d}$ matrix is usually approximated as simply
diagonal \cite{Cory2}. Accordingly, the eigenstates are then just
the basis qubits $\{|00\rangle,|01\rangle,|10\rangle,|11\rangle\}$,
so entirely lack entanglement.  That resembles our type II qubits
when $C_X = 0$, in the absence of quadrupole coupling.

Another, different sort of similarity arises from the choice of NMR
qubits as nuclear spins on different atoms within a molecule
\cite{Vandersypen}. Even for atoms of the same kind, chemical shifts
cause the effective external magnetic field to differ at different
sites. This corresponds to the role of the gradient in electric
field, emphasized in Sec.III, wherein $\triangle x > 0$ is important
both for addressing sites and for resolving the $|01\rangle$ and
$|10\rangle$ qubit pairs.

Many procedures for producing dynamical entanglement in NMR systems
by means of sequences of radiofrequency pulses have been developed
and demonstrated in performing quantum gates and algorithms
\cite{Cory,Cory2,Cory3,Vandersypen,Price,Price2}. The prospects for
adapting some of these to polar symmetric tops invite systematic
study.   We will not pursue that here, but mention an example
pertinent to resolving $\triangle\omega$, the key frequency shift
for implementing the CNOT gate.  For NMR the analog of our Eq.(28)
holds, with $\triangle\omega = J_{12}$.

Even if $\triangle\omega$ is too small to be well resolved, another
general way to perform a CNOT gate has been demonstrated in a NMR
spin system \cite{Price2}. Because qubits in both sites 1 and 2 are
in superposition states of $|0\rangle$ and $|1\rangle$, the qubit at
site 1 comprises two populations, one coupled to the qubit at site 2
in the $|0\rangle$ state and the other to the $|1\rangle$ state
there.   By means of a $\pi/2$ pulse, the qubit at site 1 can be
rotated into the transverse plane, where both populations will
undergo Larmor precession, but with different frequencies.   After a
time $\sim 1/\triangle\omega$, the two populations are $180^o$ out
of phase. Then another $\pi/2$ pulse can be performed to place both
populations at site 1 along the z-axis. The net effect is to
complete a CNOT gate with the qubit at site 2 controlling that at
site 1.  At least in principle, such procedures, well developed in
NMR, seem applicable to symmetric top type II qubit states.

\section{CONCLUSIONS AND PROSPECTS}

The seminal proposal by DeMille \cite{Demille} envisioned a quantum
computer using as qubits rotational states of ultracold polar
molecules, trapped in an optical lattice, partially oriented in an
external electric field and coupled by dipole-dipole interactions.
Many aspects and variants have been extensively studied in the
decade since, all considering diatomic molecules
\cite{Micheli,Charron,kuz,ni,lics,YelinDeMille,Wei1,Wei2}. As the
external field has an essential role, the fact that the Stark effect
is second-order for diatomic molecules has major consequences. The
field strength must be sufficiently high to induce extensive
hybridization of rotational states, so that the molecules undergo
pendular oscillations about the field direction; otherwise
rotational tumbling averages out the effective dipole moments in the
laboratory frame. As discussed in Sec. IIIC, and more fully in ref.
\cite{Wei2}, line broadening by the high field handicaps resolution
of $\triangle\omega$, the key frequency shift for 2-qubit
operations.

We find that polar symmetric top molecules offer significant
advantages.  These come primarily from the first-order Stark effect,
available for all states with $K$ and $M_J$ nonzero.   As symmetric
tops in those states precess rather than tumble, the effective
dipole moments are independent of the electric field strength
(except at high fields). Because there is no need to induce pendular
hybridization, a considerably lower external field can be used,
thereby improving prospects for resolving the $\triangle\omega$
shift. Moreover, in the first-order Stark effect the $\pm M_J$
components are readily resolved (not possible for second-order).
This enabled considering the $|J = 1, K = 1, M_J = \pm 1\rangle$
Stark components as the basis qubits (our type II), rather than
rotational states (type I). That lowers the transition frequencies
between eigenstates ({\it cf.} Table V) to the radiofrequency range,
again more congenial for resolving the $\triangle\omega$ shift. Even
more welcome, the use of $\pm M_J$ components as qubits brings forth
direct correspondences with spin-1/2 NMR systems. This opens up the
prospect of exploiting with symmetric tops a wide repertoire of
radiofrequency NMR techniques developed for quantum information
processing.

Another prospect for dealing with the small size of the
$\triangle\omega$ shift involves spatial rather than frequency
resolution.  This is exemplified by  quantum computer designs
employing superconducting flux qubits \cite{Groot}.  For these, the
generic $N = 2$ Hamiltonian in the case of transversely coupled
qubits is much like our Eqs.(18) and (19).   Instead of the Stark
terms, $\mu${\Large $\varepsilonup$} and $\mu${\Large
$\varepsilonup$}$'$, there appear single-qubit energy splittings,
denoted $\triangle_1$ and $\triangle_2$, respectively, and in place
of $\Omega_\alpha$ there appears the qubit-qubit coupling energy,
denoted by $\mathfrak{J}$ (unrelated to rotational angular momentum
or NMR spin-spin).   The analog of our $V_{d-d}$ matrix has nonzero
elements only along the anti-diagonal (equivalent to setting our
$C_0$ and $C_1 = 0$). However, for typical conditions, $\mathfrak{J}
<< (\triangle_1 - \triangle_2)$, the analog reduces just to the
simple case described under our Eq.(26) and Fig. 2; the
correspondence replaces our $C_X^2\Omega/(E_3 - E_2)$ by
$\mathfrak{J}/(\triangle_1 - \triangle_2)$. The transitions involved
in the CNOT gate (cf. Fig. 1b of ref. \cite{Groot}) then occur in
degenerate pairs, $\omega_1 = \omega_4$ and $\omega_2 = \omega_3$.
Therefore, $\triangle\omega = 0$, so frequency-selective operations
are impossible. Yet, one transition of each degenerate pair can be
selectively suppressed while coherently exciting the other, "by
simultaneously driving both qubits with the resonant frequency of
that pair, employing different amplitudes and phases" \cite{Groot}.
This method requires spatial resolution sufficient to enable qubits
on different sites to be driven individually. That may not be
feasible for our conditions, with polar molecules separated by only
0.5 $\mu$m. Such a method is well suited to a proposed design with
molecules trapped in QED cavities spaced $\sim 1$ cm (!) apart along
a superconducting transmission line resonator \cite{andre}.

As in our previous study of entanglement of polar diatomic molecules
\cite{Wei2}, we provide a generic formulation in terms of reduced
variables $(x, \triangle x, y, z, w)$.  This makes our results
applicable to a broad class of symmetric top molecules and range of
conditions envisioned for proposed quantum computers.  We also
present specific results for the $CH_3CN$ molecule \cite{Kukolich},
regarded as a particularly suitable candidate, particularly for type
II qubits. Its large dipole moment enhances the dipole-dipole
interaction and hence the $\triangle\omega$ shift, and its nitrogen
atom supplies a quadrupole moment that makes the transition dipole
$C_X$ nonzero, thereby enabling  $\triangle M_J = \pm 2$ transitions
between the type II qubits.

Many aspects important for quantum computing with polar molecules
are not discussed here (trapping operations, sources of decoherence,
and much more) because extensive analysis given for diatomic
molecules
\cite{Lee,andre,yelin,carr,Micheli,Charron,kuz,ni,lics,YelinDeMille,Wei1,Wei2}
pertains as well to symmetric tops.    We note an ironic exception.
Auxiliary storage qubits are sometimes desired to minimize
decoherence or to remove unwanted information \cite{Cory2}. Also,
"switchable dipole" schemes have been devised to in effect turn
dipole-dipole coupling "on" or "off" by transferring qubits between
states with very different dipole moments.   For diatomic molecules,
such maneuvers typically involve excited electronic states; a
prototype proposal \cite{yelin} uses CO, for which the dipole moment
in the ground $X^1\sum^+$ state is only 0.1 D, but in the metastable
excited $a^3\prod$ state is 1.5 D. For a symmetric top, such things
can be accomplished more simply by transfer to states with $K$ or
$M_J$ zero, where the first-order Stark effect vanishes. For
example, in the $J = 1$, $K = 1$ states of $CH_3CN$ under the
conditions of Table IV, for $M_J = 0$ the second-order Stark effect
\cite{Stark} yields an effective dipole moment of only 0.084 D,
whereas for $M_J = \pm 1$ the first order Stark effect gives an
effective moment of 1.96 D.   A transfer $M_J = \pm 1 \rightarrow
0$, without change in the electric field strength, would reduce the
dipole-dipole coupling 500-fold.

Symmetric tops offer many other options for qubits. Some, such as
hyperfine structure, are also available with diatomic molecules.
Others are not, such as doublet structures \cite{Weber} produced by
tunneling through barriers to inversion (e.g., in $NH_3$) or
internal rotation (e.g., in $CH_3CF_3$).  If inversion is fast
($\sim 1$ Hz for $NH_3$ in ground state), the dipole flips rapidly
and the Stark effect is second-order, whereas if inversion is slow
(e.g. $\sim 1$ year for $AsH_3$), it is first-order.    For internal
rotation involving a three-fold barrier, the tunneling doublets
occur as a nondegenerate A state, and a doubly degenerate E state;
the Stark effect for A is second- order, for E first-order.

For both diatomic and symmetric top molecules, under conditions
considered amenable for proposed quantum computers, the entanglement
of eigenstates and the associated pairwise concurrences are very
small.  Furthermore, it is not needed in the eigenstates, because
the entanglement required for computations is actually induced
dynamically.   The role of dipole-dipole coupling as the source of
eigenstate entanglement, via the off-diagonal terms of Eq.(19),
therefore is irrelevant.   Its important role is determining a
different eigenstate property, the $\triangle\omega$ shift, via
Eq.(28). The evaluation of $\triangle\omega$ does not require
eigenfunctions, only eigenvalues. This is a liberating perspective
in considering analysis of multidipole systems well beyond $N = 2$.

Mindful of the somewhat metaphysical status often accorded to
entanglement \cite{Kaiser}, we mention that fundamental theory shows
that even for symmetric tops, the "true molecular eigenstates should
not have first-order Stark effects" \cite{Klemperer2}.   That is
because the full permutation-inversion group for a molecule shows
that the only levels allowed by quantum statistics are
nondegenerate.  Yet both theory and experiment confirm that a
quasi-first-order Stark effect does appear in the presence of even a
very weak field ($<$ 0.3 V/cm) that introduces coupling between
nearly degenerate states. Hence, the very existence of first-order
Stark effect in molecules comes from field-induced entanglement.

\section*{ACKNOWLEDGEMENTS}
We are grateful for support of this work at Texas A\&M University by
the Office of Naval Research, the National Science Foundation
(CHE-0809651), and the Institute for Quantum Science and
Engineering, as well as support at Purdue by the Army Research
Office. We thank Seth Lloyd for insightful perspectives and William
Klemperer for instructive discussions of subtle aspects of molecular
dipoles.

\renewcommand{\theequation}{A\arabic{equation}}
\setcounter{equation}{0}  
\section*{APPENDIX A: QUADRUPOLE COUPLING}

As outlined in Sec.IIB, we use the "strong-field" approximation
\cite{Townes,Coester}, appropriate when the Stark shifts are much
larger than hyperfine splittings introduced by quadrupole coupling.
We need to evaluate contributions from ${\bf H}_Q$ to be added to
the qubit eigenvalues of Eqs.(10) and (13).  Also, we need to
obtain, by diagonalizing ${\bf H}_S + {\bf H}_Q$,  the modified
qubit eigenfunctions that arise from mixing of the $M_J$ Stark
components with the $M_I$ nuclear spin components.  These are
required to determine  the quadrupole contributions to the cosine
elements of Table I.    The requisite matrix elements of the ${\bf
H}_Q$ Hamiltonian,
\begin{equation}
\langle J,K,I,M_J,M_I|{\bf H}_Q|J,K,I,M'_J,M'_I\rangle
\end{equation}
are given in Eq.(33) of ref. \cite{Coester}. All contain a common
factor,
\begin{equation}
P(J,K,I) = \frac{eqQ}{4(2J-1)(2J+3)(2I-1)}\left(
\frac{3K^2}{J(J+1)}-1 \right)
\end{equation}
For the qubit states we consider,
\begin{equation}
P(1,1,1) = eqQ/40 \;\;\;\;\;\text{and}\;\;\;\;\; P(2,1,1) = -eqQ/168
\end{equation}
The elements of ${\bf H}_Q$ comprise a $9 \times 9$ matrix labeled
with $M_J = 1,0,-1$ and $M_I = 1, 0, -1$.   The first order energy
of the quadrupole hyperfine components is given by the diagonal
elements,
\begin{equation}
E_Q = P(J,K,I)\left[3M_J^2-J(J+1)\right]\left[3M_I^2-I(I+1)\right]
\end{equation}
Because the sum $M_F = M_J + M_I$ is a good quantum number, the
matrix is block diagonal, with five submatrices corresponding to
$M_F = 2, 1,0,-1,-2$ (respectively $1 \times 1$, $2 \times 2$, $3
\times 3$, $2 \times 2$, $1 \times 1$). We deal only with the $M_F =
0$ block, containing elements connecting the $(M_J, M_I) = +1, -1;
0,0; \text{and} -1,+1$ hyperfine components:
\begin{equation}
P(1,1,1)\left(\begin{array}{ccc}
1    & -3    & 6   \\
-3   &  4    & -3   \\
6    & -3    & 1
\end{array} \right)
\end{equation}
\begin{equation}
P(2,1,1)\left(\begin{array}{ccc}
-1    & -\sqrt{3}    & 6   \\
-\sqrt{3}   &  4    & -\sqrt{3}   \\
6    & -\sqrt{3}    & -1
\end{array} \right)
\end{equation}
To the diagonal elements of these matrices, we add the Stark
components, from Eq.(6), $E_S = -(\mu${\Large
$\varepsilonup$}$/J(J+1))M_J$, then carry out diagonalization to
obtain the $M_F = 0$ eigenfunctions.  As specifying $M_J$
automatically specifies $M_I$, we denote the eigenfunctions simply
by $\Psi(J, \tilde{M}_J)$, expressed as linear combinations of the
basis functions $\phi(J, M_J)$. Here we revert to wavefunction
notation, to avoid confusion with the bra notation used for qubits.
Also in labeling the eigenfunctions, we adorn $\tilde{M}_J$ with a
tilde, to indicate it is no longer a good quantum numbers because
the Stark and spin states are mixed. Performing numerical
diagonalizations led to recognition that, for $\mu${\Large
$\varepsilonup$}$\gg eqQ$, the eigenfunctions are well approximated
using for each $J$ a single mixing coefficient;  for $J = 1$:
\begin{equation}
\Psi(1,-\tilde{1}) \approx (1-a^2)\phi(1,-1)-a\phi(1,0)+a\phi(1,+1)
\end{equation}
\begin{equation}
\Psi(1,\tilde{0}) \approx a\phi(1,-1)+(1-a^2)\phi(1,0)-a\phi(1,+1)
\end{equation}
\begin{equation}
\Psi(1,+\tilde{1}) \approx -a\phi(1,-1)+a\phi(1,0)+(1-a^2)\phi(1,+1)
\end{equation}
and for $J = 2$:
\begin{equation}
\Psi(2,-\tilde{1}) \approx
(1-2b^2)\phi(2,-1)+b\phi(2,0)-\sqrt{3}b\phi(2,+1)
\end{equation}
\begin{equation}
\Psi(2,\tilde{0}) \approx -b\phi(2,-1)+(1-b^2)\phi(2,0)+b\phi(2,+1)
\end{equation}
\begin{equation}
\Psi(2,+\tilde{1}) \approx
\sqrt{3}b\phi(2,-1)-b\phi(2,0)+(1-2b^2)\phi(2,+1)
\end{equation}
The coefficients $a$ and $b$ are small positive numbers, determined
by $w = |eqQ|/\mu${\Large $\varepsilonup$}. From our numerical
results, we find
\begin{equation}
a = 0.1522w \;\;\;\;\;\text{and}\;\;\;\;\; b = 0.1789w
\end{equation}
These values are accurate within 1\% for $w < 0.1$.  Since $M_J =
\pm 1$ for our qubit states, as defined in Eqs.(9) and (12), we now
specify them  further as the hyperfine components $\Psi(J,-1)$ and
$\Psi(J,+1)$; thus for type I,
\begin{equation}
|0\rangle = |J=1,M_J=-1,M_I=+1\rangle \;\;\;\;\; \text{and}
\;\;\;\;\; |1\rangle = |J=2,M_J=-1,M_I=+1\rangle
\end{equation}
and for type II,
\begin{equation}
|0\rangle = |J=1,M_J=+1,M_I=-1\rangle \;\;\;\;\; \text{and}
\;\;\;\;\; |1\rangle = |J=1,M_J=-1,M_I=+1\rangle
\end{equation}

The quadrupole terms in the cosine elements of Table I result from
using the mixing coefficients of Eq.(A13) with Eqs.(A7) and (A10)
for type I qubits and Eqs.(A7) and (A9) for type II together with
Eq.(7) of Sec.II. In particular, for type II this gives $C_X \approx
a(1 - a^2)$.

Symmetric top molecules, other than $CH_3CN$, which contain one
nucleus with spin $I = 1$ on the symmetry axis, include:  $NH_3$ and
$NF_3$, where $^{14}N$ has $eqQ = -4.09$ MHz and 7.07 MHz,
respectively \cite{Kurolich,Sheridan}; and $CH_3D$ and $CF_3D$,
where $^2D$ has $eqQ = 191$ kHz and 171 kHz, respectively
\cite{Klemperer,Kukolich2}. In many halide molecules, such as
$CH_3X$, the halogen nuclei have $I > 1$ and large quadrupole
coupling constants \cite{Townes}. Treatment of such cases requires
use of an intermediate or weak-field approximation
\cite{Coester,Buckinghama}.

\renewcommand{\theequation}{B\arabic{equation}}
\setcounter{equation}{0}  
\section*{APPENDIX B: ENTANGLEMENT OF TWO DIPOLES}

For the ranges of reduced variables specified in Sec. III:  $x < 1$;
$10^{-4} < \triangle x < 10^{-2}$; $y < 10^{-5}$; $|z| < 5 \times
10^{-3}$; $w < 0.1$, we have obtained explicit formulas for the
coefficients of the basis qubits in Eq.(22), $\{a_i, b_i, c_i,
d_i\}$, that determine the two-dipole eigenstate entanglements.
Tables VI and VII give these formulas for types I and II qubits,
respectively. Also included are corresponding values of the pairwise
concurrence, $C_{12}$, for the eigenstates; these conform well to
the approximations of Eqs.(25) and (27). The corresponding
eigenvalues and orders-of-magnitide of the coefficients, under
conditions listed in Table IV, are in Tables II and III.
Contributions from quadrupole coupling are not included in Table VI
because these only slightly affect the entanglement for type I
qubits.  The quadrupole contributions are included in Table VII
because for type II qubits these are the sole source of eigenvalue
entanglement (since without them $C_X = 0$ and the $V_{d-d}$ matrix
of Eq.(19) is diagonal).  The quadrupole contributions enter the
entanglement coefficients in various powers of the ratio of the
quadrupole coupling to the Stark energy, $w^n$, ranging from $n = 2$
to 4.  In the concurrence values, the same dependence on $w^n$
appears.

Tables VI and VII both pertain to the regime $\triangle x \gg y$,
where the Stark shift between adjacent qubit sites is much larger
than the dipole-dipole interaction.  At present, this regime appears
most relevant for implementation prospects.  As illustrated in Fig.
2 and Eq.(27), therein the eigenfunctions differ little from the
basis qubit states, and entanglement is slight.   The extreme
opposite limit, $\triangle x = 0$, has been analyzed in ref.
\cite{Charron}; there the eigenfunctions $\Psi_2$ and $\Psi_3$
become the maximally entangled Bell states, $2^{-1/2}(|01\rangle \pm
|10\rangle)$. An interesting consequence emerged. For operation of
the CNOT gate, it was concluded that a preliminary pulse of
bandwidth much wider than the dipole-dipole interaction should be
applied. It would entirely undo the entanglement by forming $\pm$
combinations of the Bell states and thereby unwed the $|01\rangle$
and $|10\rangle$ qubits.

\begin{table}[H]
\begin{minipage}{16cm}
\centering \caption{Eigenfunction entanglement coefficients, type I
qubits$^a$}
\begin{tabular}{c}
\hline\hline
$\Psi_i = a_i|00\rangle + b_i|01\rangle + c_i|10\rangle + d_i|11\rangle$ \\
\hline\let\thefootnote\relax\footnotetext{$^a$Coefficients
\{$a_i,b_i,c_i,d_i$\} of eigenfunctions $i=1 \rightarrow 4$ obtained
from numerical diagonalization of the matrices of Eqs(18) and (19).
Table II gives the corresponding eigenvalues as well as
orders-of-magnitude of the coefficients under conditions listed in
table II. Values are included for the pairwise concurrence,
$C_{12}$, and conform well to the approximations of Eqs.(25) and
(27).}
$a_1 = (0.07y)^2/2$; \;\;\;\;\;\;\;\;\;\;\;\; $b_1 = c_1 = (0.048+0.0044x)y$; \;\;\;\;\;\;\;\;\;\;\;\;\;\;\;\; \\
$d_1 = -a_4 = (0.019+0.0017x)y$; \;\;\;\; $C_{12} = K(x,x')y \approx 2d_1$;  \\

$a_2 = -(0.028+0.0026x)y$; \;\;\;\;\; $b_2 = 1 - c_2^2/2$; \;\;\;\;\; $c_2 = -0.454y/\triangle x$; \\
$d_2 = (0.009+0.0009x)y$; \;\;\;\; $C_{12} \approx 2|c_2|$;  \\

$a_3 = -(0.062+0.0058x)y$; \;\;\;\;\;\;\; $b_3 = - c_2 = 0.454y/\triangle x$; \;\;\;\; $c_3 = b_2$; \\
$d_3 = (0.021+0.0019x)y$; \;\;\;\; $C_{12} \approx 2|c_2|$;  \\

$a_4 = -d_1 = (0.019+0.0017x)y$; \;\;\;\; $b_4 = c_4 = -(0.016+0.0015x)y$;  \\
$d_4 = -a_4 = 1 - (0.04y)^2/2$; \;\;\;\; $C_{12} = K(x,x')y \approx 2a_4$;  \\

\hline \hline
\end{tabular}
\end{minipage} \label{table-typeI-all}
\end{table}

\begin{table}[H]
\begin{minipage}{16cm}
\centering \caption{Eigenfunction entanglement coefficients, type II
qubits$^a$}
\begin{tabular}{c}
\hline\hline
$\Psi_i = a_i|00\rangle + b_i|01\rangle + c_i|10\rangle + d_i|11\rangle$ \\
\hline\let\thefootnote\relax\footnotetext{$^a$Footnote to Table IV
pertains have as well, except that corresponding eigenvalues and
order-of-magnitude values are in Table III. Contributions from
quadrupole couplings are included with $w = |z|/x =
|eqQ|/\mu${\Large $\varepsilonup$} and $z = eqQ/B$}
$a_1 = 1 - b_1^2$; \;\;\;\;\;\;\;\;\;\;\;\;\;\;\;\;\;\;\;\;\;\;\;\;\;\;\;\;\; $b_1 = c_1 = -0.0786w^2y/z$; \;\;\;\;\;\;\;\;\;\;\;\;\;\;\;\;\; \\
$d_1 = -0.0118w^3y/z$; \;\;\;\;\;\;\;\;\;\; $C_{12} = 0.0247w^3y/z \approx 2d_1$;  \\

$a_2 = d_2 = a_3 = d_3 = 0.0743w^2y/z$; \;\;\;\;\;\; $b_2 = -c_3 = -0.0225w^2y/\triangle x$; \;\;\;\;  \\
$c_2 = 1 - a_2^2$; \;\;\;\; $d_2 = a_2$; \;\;\;\; $C_{12} = 0.044w^2y/\triangle x \approx 2|b_2|$;  \\

$a_3 = d_3 = a_2 = d_2$; \;\;\;\; $b_3 = c_2 = 1 - a_2^2$; \;\;\;\; $c_3 = -b_2 = 0.0225w^2y/\triangle x$; \\
$d_3 = a_3$; \;\;\;\;\;\;\;\;\;\;\;\;\;\;\;\;    $C_{12} = 0.044w^2y/\triangle x \approx 2|c_3|$; \;\;\;\;\; \\

$a_4 = 0.0107w^3y/z$; \;\;\;\;\;\;\;\;\;\;\;\;\;\;\;\;\;\;\; $b_4 = c_4 = -0.0715w^2y/z$; \;\;\;\;\;\;\;\;\;\;\;\;\;\;\;\; \\
$d_4 = 1 - b_4^2$; \;\;\;\;\;\;\;\;\;\;\;\;\;\;\;\; $C_{12} = 0.0205w^3y/z \approx 2a_4$; \;\; \\

\hline \hline
\end{tabular}
\end{minipage} \label{table-typeII-all}
\end{table}

\newpage

\newpage
\begin{figure}[t]
\begin{center}
\includegraphics[width=0.75\textwidth]{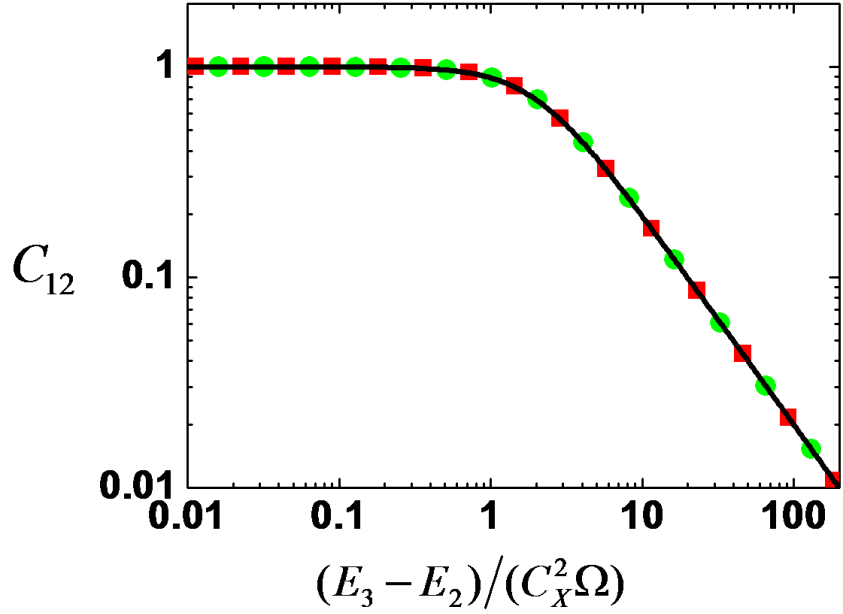}
\end{center}
\caption{(Color online)  Pairwise concurrence $C_{12}$ for
eigenstates 2 and 3 of two symmetric top dipoles entangled via
dipole-dipole interaction, as a function of the ratio of the
difference of the eigenvalues, $(E_3 - E_2)$, to the element,
$C_X^2$ that connects the $|01\rangle$ and $|10\rangle$ basis qubits
in the $V_{d-d}$ matrix of Eq.(19).    The difference $(E_3 - E_2)/B
= \triangle x/3$ and $\triangle x$, for type I and type II qubits,
respectively, as seen in Tables II and III.   Points (green for type
I, red for II) were obtained from numerical calculations including
all elements of the $V_{d-d}$ matrix; curve (black) from the
minimalist $2 \times 2$ model of Eqs.(25-27). The same $C_{12}$
function applies to spin-1/2 NMR systems, with $E_3 - E_2 =
g\mu_N(H' - H)$ and $C_X^2\Omega_\alpha$ replaced by
$\frac{1}{2}J_{12}$, where $g$ is the nuclear $g$-factor, $\mu_N$
the nuclear magneton, $H$ the magnetic field strength, and $J_{12}$
the spin-spin coupling constant.} \label{Fig2}
\end{figure}

\newpage
\begin{figure}[t]
\begin{center}
\includegraphics[width=0.60\textwidth]{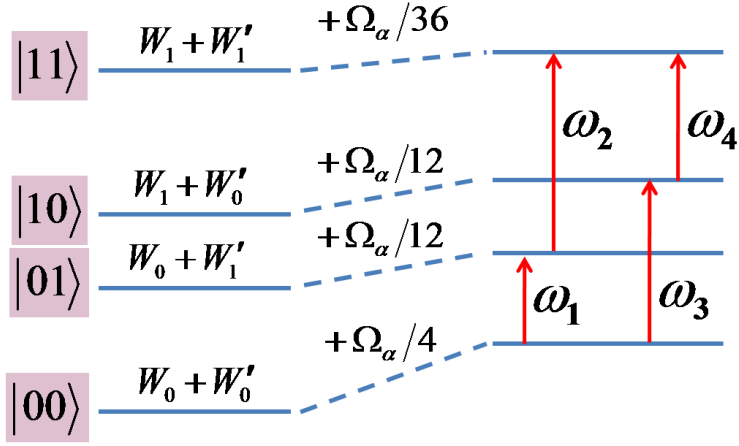}
\end{center}
\caption{(Color online)  Schematic energy levels for type I qubit
eigenstates of two symmetric top dipoles.  At left are indicated
qubit basis states, with corresponding eigenenergies from Eqs.(18)
and (19).  Contributions from quadrupole coupling are not shown (but
included in Tables II and V).  At right are transitions that are
involved in CNOT operation: $\omega_1$ transfers the dipole at site
2 from $|0\rangle$ to $|1\rangle$, with dipole at site 1 remaining
in $|0\rangle$,  then  2 transfers dipole at site 1 from $|0\rangle$
to $|1\rangle$ with dipole at site 2 remaining in $|1\rangle$. The
same result could be reached by $\omega_3$ followed by $\omega_4$.
Transition frequencies are given in Table V.} \label{Fig3}
\end{figure}

\newpage
\begin{figure}[t]
\begin{center}
\includegraphics[width=0.60\textwidth]{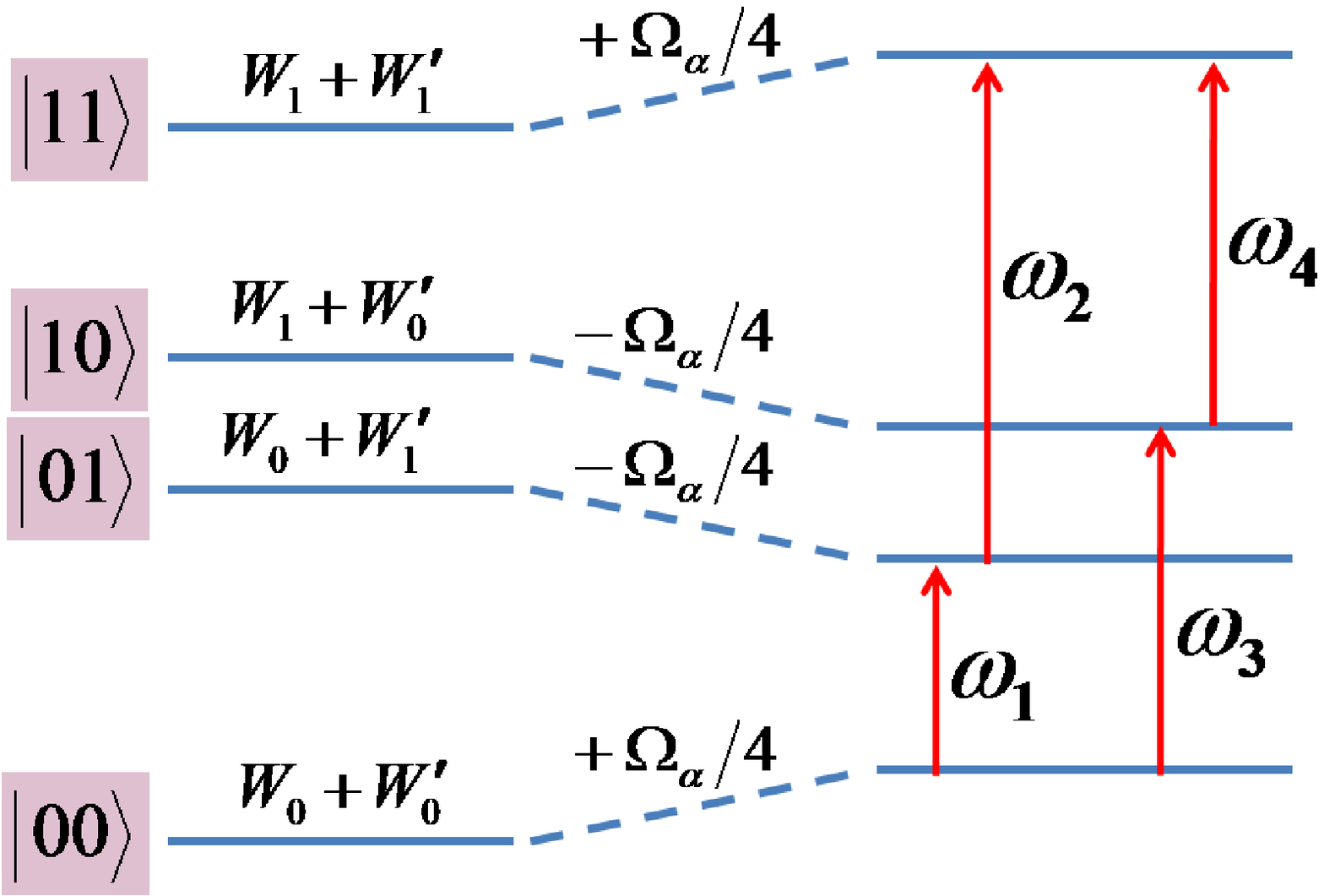}
\end{center}
\caption{(Color online)   Schematic energy levels for type II qubit
eigenstates of two symmetric top dipoles; format as in Fig. 3.
Eigenenergies, including quadrupole coupling are (not shown) are
given in Tables III and transition frequencies in Table V.}
\label{Fig4}
\end{figure}


\begin{thebibliography}{71}

\bibitem{Bennett} C. H. Bennett, Int. J. Theor. Phys. {\bf 21}, 905 (1982)

\bibitem{Deutsch} D. Deutsch, Proc. R. Soc. London Ser. A {\bf 400}, 97 (1985)

\bibitem{Feynman} R. P. Feynman, Found. Phys. {\bf 16}, 507 (1986)

\bibitem{Shor} P. W. Shor, \textit{Proceedings of the 35th Annual Symposium on Foundations of Computer Science}, edited by S. Goldwater, (IEEE Computer Society Press, Los Alamitos, CA, 1994)

\bibitem{Grover} L. K. Grover, Phys. Rev. Lett. {\bf 79}, 325 (1997)

\bibitem{Chuang} I. L. Chuang, N. Gershenfeld and M. Kubinec, Phys. Rev. Lett. {\bf 80}, 3408 (1998)

\bibitem{Seth} S. Lloyd, Science {\bf 261}, 1569 (1993)

\bibitem{DiVincenzo} D. P. DiVincenzo, Science {\bf 270}, 255 (1995)

\bibitem{Gershenfeld} N. A. Gershenfeld, and I. L. Chuang, Science {\bf 275}, 350 (1997)

\bibitem{Cory} D. G. Cory, A. F. Fahmy and T. F. Havel, Proc. Natl. Acad. Sci. {\bf 94}, 1634 (1997)

\bibitem{Kane} B. E. Kane, Nature {\bf 393}, 133 (1998)

\bibitem{Loss} D. Loss and D. P. DiVincenzo, Phys. Rev. A {\bf 57}, 120 (1998)

\bibitem{wallraff} A. Wallraff, D. I. Schuster, A. Blais, L. Frunzio,R. S. Huang, J. Majer, S. Kumar, S. M. Girvin and R. J. Schoelkopf, Nature {\bf 431}, 162 (2004).

\bibitem{Lee} C. Lee and E. A. Ostrovskaya, Phys. Rev. A {\bf 72}, 062321 (2005).

\bibitem{Demille} D. DeMille, Phys. Rev. Lett. {\bf 88}, 067901 (2002).

\bibitem{andre} A. Andre, D. DeMille, J. M. Doyle, M. D. Lukin, S. E. Maxwell, P. Rabl, R. J. Schoelkopf and P. Zoller, Nature Phys. {\bf 2}, 636 (2006)

\bibitem{yelin} S. F. Yelin, K. Kirby and R. Cote, Phys. Rev. A {\bf 74}, 050301(R) (2006).

\bibitem{carr} L. D. Carr, D. DeMille, R. V. Krems and J. Ye, New J. Phys. {\bf 11}, 055049 (Focus Issue) (2009).

\bibitem{Book2009} R. V. Krems, W. C. Stwalley and B. Friedrich, Eds.
\textit{Cold molecules: theory, experiment, applications} (Taylor
and Francis, 2009).

\bibitem{Friedrich} B. Friedrich and J. M. Doyle, ChemPhysChem {\bf 10}, 604 (2009).

\bibitem{Kotochigova} S. Kotochigova and E. Tiesinga, Phys. Rev. A {\bf 73}, 041405(R) (2006).

\bibitem{Micheli} A. Micheli, G. K. Brennen and P. Zoller, Nature Phys. {\bf 2}, 341-347 (2006).

\bibitem{Charron} E. Charron, P. Milman, A. Keller and O. Atabek, Phys. Rev. A {\bf
75}, 033414 (2007); Erratum, Phys. Rev. A {\bf 77}, 039907 (2008).

\bibitem{kuz} E. Kuznetsova, R. Cote, K. Kirby and S. F. Yelin, Phys. Rev. A {\bf 78}, 012313 (2008).

\bibitem{ni} K. K. Ni, S. Ospelkaus, M. H. G. de Miranda, A. Peer, B. Neyenhuis, J. J. Zirbel, S. Kotochigova, P. S. Julienne, D. S. Jin and J. Ye,
Science {\bf 322}, 231 (2008).

\bibitem{lics} J. Deiglmayr, A. Grochola, M. Repp, K. Mortlbauer, C. Gluck, J. Lange, O. Dulieu, R. Wester and M.
Weidemuller, Phys. Rev. Lett. {\bf 101}, 133004 (2008).

\bibitem{YelinDeMille} S. F. Yelin, D. DeMille and R. Cote, \textit{Quantum information processing
with ultracold polar molecules} in \cite{Book2009}, p. 629 (2009).

\bibitem{Wei1} Q. Wei, S. Kais and Y. Chen, J. Chem. Phys. {\bf 132}, 121104 (2010).

\bibitem{Wei2} Q. Wei, S. Kais, B. Friedrich and D. Herschbach, J. Chem. Phys. {\bf 134}, 124107 (2011).

\bibitem{Cory2} D. G. Cory, R. Laflamme, E. Knill, L. Viola, T. F. Havel,
N. Boulant, G. Boutis, E. Fortunato, S. Llloyd, R. Martinez, C.
Negrevergne, M. Pravia, Y. Sharf, G. Teklemariam, Y. S. Weinstein
and W. H. Zurek, Fortschr. Phys. {\bf 48}, 875 (2000)

\bibitem{Lieven} L. M. K. Vandersypen, C. S. Yannoni and I. L. Chuang, Liquid State NMR Quantum Computing,
\textit{Encyclopedia of Nuclear Magnetic Resonance, Volume 9:
Advances in NMR} (Edited by David M. Grant and Robin K. Harris)
(John Wiley \& Sons, Ltd, Chichester, 2002)

\bibitem{Townes} C. H. Townes and A. L. Schawlow, \textit{Microwave Spectroscopy} (McGraw-Hill, New York, 1955)

\bibitem{Zare} R. N. Zare, \textit{Angular Momentum} (John Wiley \& Sons, USA, 1988)

\bibitem{Stark} For comparison, for $J = 1$ states with $K = 0$ or
1 and $M_J = 0$ or 1, the second-order Stark energy is
$[(\mu${\Large $\varepsilonup$}$)^2/20B](2-3M_J^2)$, and the
effective dipole moment is $(\mu^2${\Large
$\varepsilonup$}$/10B)(2-3M_J^2)$.

\bibitem{Meerakker} S. Y. T. van de Meerakker, H. L. Bethlem, and G. Meijer, \textit{Slowing, Trapping, and Storing of Polar Molecules by Means of Electric Fields} in \cite{Book2009}, pp 509-552 (2009).

\bibitem{Klemperer} S. C. Wofsy, J. S. Muenter and W. Klemperer, J. Chem. Phys. {\bf 53}, 4005 (1970)

\bibitem{Coester} F. Coester, Phys. Rev. {\bf 77}, 454 (1950)

\bibitem{Kukolich} S. G. Kukolich, J. Chem. Phys. {\bf 76}, 97 (1982)

\bibitem{Wootters} W. K. Wootters, Phys. Rev. Lett. {\bf 80}, 2245 (1998).

\bibitem{Thanks} Our discussion of the CNOT operation is largely drawn from tutorial instruction kindly given us by D. DeMille, amplifying the discussion of Fig.17.2 in ref. \cite{wallraff}.

\bibitem{Jones2} J. A. Jones and M. Mosca, J. Chem. Phys. {\bf 109}, 1648 (1998)

\bibitem{Shioya} K. Shioya, K. Mishima and K. Yamashita, Mol. Phys. {\bf 105}, 1283 (2007)

\bibitem{Mishima} K. Mishima and K. Yamashita, Chem. Phys. {\bf 361}, 106 (2009)

\bibitem{Chen} J. Chen, C. Li, C. Hwang and Y. Ho, J. Chem. Phys. {\bf 134}, 134103 (2011)

\bibitem{Reina} J. H. Reina, R. G. Beausoleil, T. P. Spiller and W. J. Munro, Phys. Rev. Lett. {\bf 93} 250501 (2004).

\bibitem{Sugny} D. Sugny, L. Bomble, T. Ribeyre, O. Dulieu, and M. Desouter-Lecomte, Phys. Rev. A {\bf 80}, 042325 (2009)

\bibitem{Barker} P. F. Barker and S. M. Purcell and M. N. Shneider, Phys. Rev. A {\bf 77}, 063409 (2008)

\bibitem{Zelevinsky} T. Zelevinsky, S. Blatt, M. Boyd, G. Campbell, A. Ludlow and J. Ye,, ChemPhysChem {\bf 9}, 375 (2008)

\bibitem{Vandersypen} L. M. K. Vandersypen, M. Steffen, G. Breyta, C. S. Yannoni, M. H. Sherwood and I. L. Chuang, Nature {\bf 414}, 883 (2001).

\bibitem{Cory3} D. G. Cory, M. D. Price, T. F. Havel, Phys. D {\bf 120}, 82 (1998)

\bibitem{Price} M. D. Price, S. S. Somaroo, A. E. Dunlop, T. F. Havel and D. G.
Cory, Phys. Rev. A {\bf 60}, 2777 (1999)

\bibitem{Price2} M. D. Price, S. S. Somaroo, C. H. Tseng, J. C. Gore, A. F. Fahmy,T. F. Havel and D. G. Cory, J. Magn. Reson. {\bf 140}, 371 (1999)

\bibitem{Groot} P. C. de Groot, J. Lisenfeld, R. N. Schouten, S. Ashhab, A. Lupascu,
C. J. P. M. Harmans, and J. E. Mooij, Nature Phys. {\bf 6}, 763
(2010)

\bibitem{Weber} W. H. Weber and R. W. Terhune, J. Phys. Chem. {\bf 78}, 6437 (1983)

\bibitem{Kaiser} D. Kaiser, \textit{How the Hippies Saved Physics} (WW.Norton, New York, 2011).

\bibitem{Klemperer2} W. Klemperer, K. K. Lehmann, J. K. G. Watson and S. C. Wofsey, J. Phys. Chem. {\bf 97}, 2413 (1993)

\bibitem{Kurolich} S. G. Kurolich and S. C. Wofsy, J. Chem. Phys. {\bf 52}, 5477 (1970)

\bibitem{Sheridan} J. Sheridan and W. Gordy, Phys. Rev. {\bf 79}, 513 (1950)

\bibitem{Kukolich2} S. G. Kukolich, A. C. Nelson and D. J. Ruben, J. Mol. Spec. {\bf 40}, 33 (1971)

\bibitem{Buckinghama} A. D. Buckinghama and P. J. Stephensa, Mol. Phys. {\bf 7}, 481 (1964)


\end{thebibliography}
\end{document}